%% file: main.tex
\documentclass[acmtog]{acmart}

\usepackage{algcompatible}
\usepackage{subcaption}
\usepackage{xspace}
\usepackage{booktabs}
\usepackage{graphicx}
\usepackage{todonotes}
\usepackage{soul}
\usepackage{algorithm}
\usepackage[noend]{algpseudocode}

\DeclareRobustCommand{\fmeasure}{$\mathcal{F}-$measure\xspace}
\DeclareRobustCommand{\Otsuka}{\textit{Otsuka$14$}\xspace}
\DeclareRobustCommand{\Taylor}{\textit{Taylor$15$}\xspace}
\DeclareRobustCommand{\Cho}{\textit{Cho$16$}\xspace}
\DeclareRobustCommand{\Gmiden}{\textit{Gmiden$16$}\xspace}
\DeclareRobustCommand{\Song}{\textit{Song$16$}\xspace}
\DeclareRobustCommand{\Moore}{\textit{Moore$17$}\xspace}
\DeclareRobustCommand{\Stabili}{\textit{Stabili$19$}\xspace}
\DeclareRobustCommand{\Olufowobi}{\textit{Olufowobi$20$}\xspace}

\AtBeginDocument{%
  \providecommand\BibTeX{{%
    \normalfont B\kern-0.5em{\scshape i\kern-0.25em b}\kern-0.8em\TeX}}}

\acmJournal{TCPS}

\begin{document}
\title{Performance comparison of timing-based anomaly detectors for Controller Area Network: a reproducible study}

\author{Francesco Pollicino}
\authornote{Both authors contributed equally to this research.}
\email{francesco.pollicino@unimore.it}
\orcid{0000-0002-2421-1852}
\author{Dario Stabili}
\authornotemark[1]
\email{dario.stabili@unimore.it}
\orcid{0000-0001-6850-334X}
\author{Mirco Marchetti}
\email{mirco.marchetti@unimore.it}
\orcid{0000-0002-7408-6906}
\affiliation{%
  \institution{University of Modena and Reggio Emilia}
  \streetaddress{Via P. Vivarelli, 10}
  \city{Modena}
  \country{ITA}
}



\renewcommand{\shortauthors}{Pollicino, Stabili, and Marchetti}

\input{sections/abstract.tex}



\maketitle

\input{sections/introduction.tex}

\input{sections/primer.tex}
\input{sections/related.tex}
\input{sections/dataset.tex}
\input{sections/implementations.tex}

\input{sections/experimental_evaluation.tex}
\input{sections/conclusions.tex}

\bibliographystyle{ACM-Reference-Format}
\bibliography{sections/bibliography}

\end{document}

%% file: sections/abstract.tex
\begin{abstract}
This work presents an experimental evaluation of the detection performance of eight different algorithms for anomaly detection on the Controller Area Network (CAN) bus of modern vehicles based on the analysis of the timing or frequency of CAN messages. 
This work solves the current limitations of related scientific literature, that is based on private dataset, lacks of open implementations, and detailed description of the detection algorithms. These drawback prevent the reproducibility of published results, and makes it impossible to compare a novel proposal against related work, thus hindering the advancement of science. This paper solves these issues by publicly releasing implementations, labeled datasets and by describing an unbiased experimental comparisons.


\end{abstract}

%% file: sections/introduction.tex
\section{Introduction}
\label{s:introduction}

Automotive Cyber Security is a relatively new research area, that is rapidly growing to encompass many different security issues: from the design of security countermeasures aiming to deter cyber attackers, to detection mechanisms trying to identify ongoing attacks and reactive countermeasures to contain and respond to malicious activities. One of the most active research areas in this field is the design of novel intrusion detection systems applied to the Controller Area Network (CAN) bus, one of the prominent network technologies used to interconnect Electronic Control Units (ECUs) deployed within modern vehicles. Several intrusion detection algorithms have already been proposed~\cite{Dupont2019Survey}, and it is possible to classify them based on the features of the CAN communication that they use to identify anomalies and attacks. 

This paper focuses on the subset of CAN intrusion detection systems that identifies anomalies by analyzing the timing with which CAN messages are sent over the CAN bus. This detection approach is promising, since many powerful attacks (such as fuzzing~\cite{Lee2015Fuzzing}, injection~\cite{Miller2015Exploitation} and Denial of Service~\cite{Bozdal2018Denial}) are based on the injection of malicious messages on the CAN bus in addition to the legitimate traffic. Our main goal is to propose an unbiased comparison of the detection performance of time-based anomaly detectors for the CAN bus. We stress that this task is much more complex than it might seem. A direct comparison among different detection algorithms published in the scientific literature is hindered by many different issues. Different papers use different detection metrics to evaluate the performance of the proposed solution. Each work uses a proprietary dataset for carrying out the experimental performance evaluation, and in most of the cases the dataset is not publicly released, it is only described at a very high level, and lacks all the details required to replicate the attacks. Moreover, novel solutions designed to detect anomalies in CAN communication are not compared to existing solutions, or are compared with naive detection metrics to demonstrate that a more complex solution is better in the detection task. In the vast majority of cases the authors of a scientific paper do not disclose a reference implementation, thus requiring other researchers to re-implement the proposed algorithm. More often than not, the paper only includes a high level description of the proposed algorithm that lacks many relevant details that are actually required for a real implementation. Finally, several papers omit important aspects related to the tuning and training of the proposed algorithm that have a strong impact on their detection performance.

This work tackles three of the important limits in this field of research. The first major contribution is the empirical and unbiased comparison of eight different time-based CAN anomaly detectors over two different datasets, an original one and one that is already publicly available. The second contribution to the state-of-the-art is to foster the reproducibility of similar studies. We publicly release all the reference implementation of the detectors considered in this study, together with the novel dataset used to tune and test the detection algorithm. This contribution allows all researchers and industry practitioners to fully replicate our research results, validate the correctness of our reference implementations, assess the quality of the dataset, and easily compare a novel proposal with respect to the state-of-the-art. Finally, we highlight the limits of publicly available datasets used for the experimental evaluation of existing solutions against the dataset presented at first in this work. This analysis also demonstrate that it is crucial to identify a comprehensive threat model to demonstrate how anomaly detectors are effected by attacks, such as the effects of different cycle times and injection frequency on the overall performance evaluation.

The remainder of the paper is structured as follows. Section~\ref{s:primer_can} discusses the main characteristics of the CAN bus and of CAN messages that are required to understand this work. Section~\ref{s:related} presents the related work and describes in detail the eight detectors considered in this paper. Section~\ref{s:dataset} describes the dataset used throughout the paper to validate and test the detection algorithms. Section~\ref{s:implementations} presents our reference implementation of the eight detectors, focusing on the additional assumption and design choices that are missing in the original papers. Section~\ref{s:experimental_evaluation} compares the detection performance of our reference implementations against several attack instances, involving different CAN messages and injection frequencies. Finally, Section~\ref{s:conclusions} concludes the paper.

%% file: sections/primer.tex
\section{A primer on the Controller Area Network}
\label{s:primer_can}

The Controller Area Network (CAN) is one of the communication protocols used between the Electronic Control Units (ECU) deployed within the vehicle~\cite{BoschCANSpec}. The CAN bus is designed to enable communication between the nodes without requiring a host computer. CAN is one of the most deployed networking protocols for in-vehicular communications due to its high resilience to electromagnetic interferences and its low implementation costs. 
It is a broadcast-message based communication protocol, and transmission on the CAN bus uses a bit-wise arbitration method for contention resolution. When two different nodes start transmission of a frame at the same time, the node with the highest priority continues sending the frame without interruption, while the other node backs off and re-tries transmission at a later time.

\begin{figure}
    \centering
    \includegraphics[width=0.9\columnwidth]{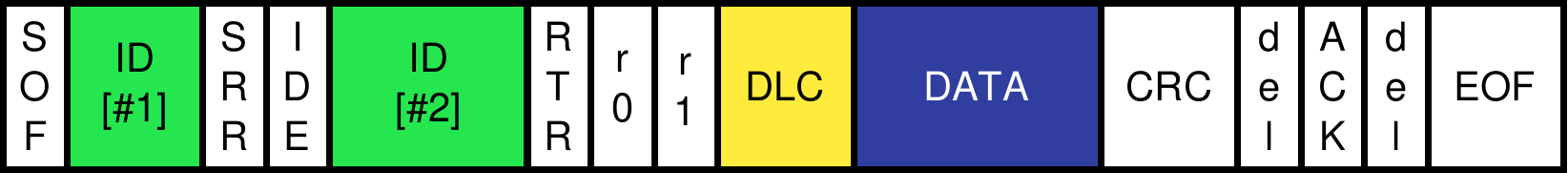}
    \caption{CAN data frame in the extended format}\label{fig:data_frame_extended}
\end{figure}

The CAN protocol defines $4$ different types of frames with different usages, but only the \textit{data frame} is used to transmit data between ECUs. 
The main fields composing the CAN data frame are the \emph{identifier (ID)}, the \emph{data length code (DLC)}, and the \emph{payload (data)}. The \emph{ID} identifies the CAN data frame and the content of its \emph{data} field. Each ECU transmits only a limited set of messages with a particular ID while receiving ECUs use the value of the ID field to select data frames relevant for their functioning. A message with a particular value of the ID field is always sent by only one ECU.
The ID field is also used for arbitration of the CAN messages, where lower values of this field denote messages with higher priority. In the current standard for basic CAN communication, the ID field is defined with a size of either $11$ (standard format) or $29$ bits (extended format). Figure~\ref{fig:data_frame_extended} shows the structure of an extended CAN message. Note that the extra $18$ bits of the \emph{extended} format (ID \#2) are encoded separately from the $11$ bits of the \emph{standard} format (ID \#1) for backward compatibility. 

The \emph{DLC} field encodes the number of bytes composing the \textit{data} field, and has a size of $4$ bits. Since the maximum length of the \textit{data} field is $8$ bytes, valid DLC values ranges from $0$ to $8$, while values from $9$ to $15$ are left unused.

The \emph{data} field encapsulates the information that the sender ECU transmits to other ECUs on the network. The data field has a variable size (from $0$ to $8$ bytes) and usually packs several different signals. The CAN standard leaves complete freedom to the car manufacturers about the structure, number, encoding, and semantic of these signals. Hence, without having access to the formal specifications of the CAN messages for a particular vehicle model, the data encoded in the data field can only be interpreted as an opaque binary blob.

Between two consecutive CAN data frames, an \emph{interframe space} is required. The \emph{interframe space} consists of at least three consecutive recessive bits, called \emph{interframe bits}. Following the interframe space, if a dominant bit is detected, it is considered as the \emph{start-of-frame} of the next data frame.

%% file: sections/related.tex
\section{Related work}
\label{s:related}
Since the introduction of microcontrollers to in-vehicle networks, the automotive domain is considered one of the most prominent examples of Cyber-Physical Systems (CPSs). 


The main characteristic of an automotive CPS is the central role of the automotive in-vehicle network, which is used by the ECUs to exchange data for their operational needs. With the increasing development of advanced driver assistance systems (ADAS), cyber-security researchers started to demonstrate attacks to these features. Miller and Valasek~\cite{Miller2014Adventures,Miller2015Exploitation} demonstrated the consequences of an Internet-based, remote cyber-attack to a modern, unmodified, licensed vehicle, with a huge media coverage. Since then, many researchers started to develop Intrusion Detection Systems (IDS) by applying concepts borrowed from classical computer networks to the in-vehicle networks~\cite{Dupont2019Survey, Hyo2021Survey}. 
Some of these solutions focus on the analysis of the low-level characteristics of the ECUs~\cite{Cho2017Viden, Kneib2018Scission}, other solutions focus on the analysis of the in-vehicle network communications. Of this last group of solutions, security researchers have developed intrusion detection systems based on the statistical analysis of the content of the CAN bus~\cite{Muter2011Entropy, Marchetti2016Entropy, nowdehi2019casad}, while other solutions are focused on the analysis of the content of the CAN data frames~\cite{Ling2012Unknown,Marchetti2017Sequences, Stabili2021DAGA, Stabili2017Hamming,Tomlinson2021Context}.

Most of these works are only focused on a particular aspect of CAN communication, thus increasing the complexity of comparing novel solutions with existing literature. Moreover, only a handful implementations of the current state-of-the-art is publicly available, preventing the comparison of a novel solution with the existing ones.

The scope of this work is to tackle this problem by proposing an unbiased comparison of existing solutions designed to detect anomalies on CAN communication using the timings of the CAN messages as their detection feature. We chose this group of solutions since they represent one of the most analyzed group of solutions for the anomaly detection task on CAN communications, and they can be deployed as a software-only solution, without requiring dedicated hardware components to their functioning.
The first work focused on this aspect is presented in~\cite{Otsuka2014CANSecurity}, while following works are focused on either frequency~\cite{Taylor2015Frequency, Moore2017Frequency} or timing analysis~\cite{Song2016Intervals, Gmiden2016Detection, Stabili2019Missing}. All these works are based on the assumption that most CAN messages are sent periodically on the network within a fixed time interval, hence it is possible to exploit this feature to detect messages that do not follow the expected timing. However, these detection methods are only applicable to cyclic messages and cannot detect any anomaly if the attack targets a non cyclic message.

In the following sections, we describe the anomaly detectors considered for the experimental evaluation presented in this work. For readability purpose, we associate each algorithm with a label composed by the last name of the first author and the last two digits of the publication year. For clarity, we uniformed the names of the main parameters, variables, and attack scenarios described in the original works.

\subsection{Otsuka14}
In the work presented in~\cite{Otsuka2014CANSecurity} the authors propose an anomaly detection algorithm that uses a delayed-decision cycle detection method to detect (and possibly prevent) spoofing attacks. 
The algorithm presented in \Otsuka assumes that, since data frames are transmitted at a constant cycle time $ct$, any modification of the normal behavior of CAN communication should change its cycle time. 
When an ECU receives a data frame with the same CAN ID of the previous data frame and a cycle time less than $ct + \delta$ (where $\delta$ is a threshold parameter), the ECU holds the data frame until the expected time $T$ is passed. In case another message with the same ID is received in the waiting period, then the ECU is able to detect an ongoing attack and the data frames are not processed. 
The algorithm presented in \Otsuka is trained with real CAN data and tested against the injection of two different message IDs, one exhibiting a stable cycle time $ct$ while the other exhibits a non-cyclic pattern.  

The results are presented by means of False Positive Rate (FPR) and False Negative Rate (FNR). Despite the comparison of the detection performance with the previous state-of-the-art is not discussed, we remark that \Otsuka is, to the best of our knowledge, the first paper presenting a detection algorithm based on the timing analysis of CAN communications.
The computational overhead of the detection algorithm \Otsuka is equal to $\mathcal{O}(1)$ for each received CAN message.

\subsection{Taylor15}
In the work presented in~\cite{Taylor2015Frequency}, the authors design an anomaly detection algorithm for the detection of anomalies based on inter-packet time over a sliding window. In particular, the algorithm presented in \Taylor uses test values over consecutive CAN flows (defined as a sequence of CAN data frames) for its detection task. Each test value is evaluated as a $t-$test, comparing the mean time difference with its historical value (i.e. the cycle time $ct$ of the same CAN ID). The algorithm then uses the test values for the evaluation of the anomaly score (defined as a logarithmic sum) over a sequence of scores, to identify anomalies in the CAN communication. 
This algorithm is tested against the injection and the removal of CAN messages with different attack duration, ranging from $100$ milliseconds to $1$ second. For the message injection attack and for each duration, the injected packed is inserted once, five, and ten times its normal frequency. The detection performance are presented by means of the Receiver Operating Characteristics (ROC) and the Area Under Curve (AUC) measure. The authors do not compare the algorithm with previous works but only with a One-Class Support Vector Machine classifier trained on the same data. 
The computational overhead of the detection algorithm \Taylor is equal to $\mathcal{O}(1)$ for each received CAN message.

\subsection{Cho16}
\label{ss:Cho16}
The authors of~\cite{Cho2016Fingerprinting} design and test a \emph{Clock-based Intrusion Detection System} (CIDS) for CAN communications. This algorithm (labeled as \Cho) leverages the intervals of periodic in-vehicle messages for fingerprinting ECUs. Then, the fingerprints are used for constructing a baseline of ECUs' clock behaviors based on the Recursive Least Squares (RLS) algorithm. The Cumulative Sum (CUMSUM) is then computed on this baseline to detect anomalies in message timings thanks to an adaptive threshold. The \Cho algorithm  is able to detect anomalies inside a window of size $W$ dataframes, and does not identify a single dataframe as malicious. 
This algorithm is presented in two versions: the \emph{per-message} detection and the \emph{message-pairwise} detection. The latter version (\emph{message-pairwise detection}) is based on the assumption that the number and identity of CAN messages generated from the same ECU are known. We remark that these information can only be known a-priori by either having access to the DBC or by apply modern mapping techniques such as~\cite{Kulandaivel2019CANVas}. Since we only relies on CAN log traces and we do not have access to the DBC of our test vehicle, only the \emph{per-message detection} version can be applied to our datasets. 
The \Cho algorithm is tested against the injection of CAN messages (called \emph{fabrication attack} in the original work) on both a CAN bus prototype and a real vehicle. On the real vehicle scenario, authors target a message with a cycle time $ct$ equals to $20ms$, however the original paper does not contain any information about the injection frequency. Moreover, the paper does not include performance comparison against previous work. 
The computational overhead of the detection algorithm \Cho is equal to $\mathcal{O}(N^2)$ for each window, where $N$ is the size of the data matrix. We remark that this computational complexity is equal to the one of the \emph{RLS algorithm} used for constructing the baseline of ECUs' clock behaviors.

\subsection{Gmiden16}
\label{ss:Gmiden16}
In the work presented in~\cite{Gmiden2016Detection}, the authors design a lightweight intrusion detection algorithm based on the analysis of the frequencies of the CAN messages.
In particular, the algorithm presented in \Gmiden uses the frequency of a CAN message to detect anomalies in the CAN communication. Upon reception of a message, the algorithm compares the time difference $\Delta_{t}$ between the new message with the previous one sharing the same ID, and generates an anomaly in case the time difference is less than half the estimated cycle time $ct$.
In~\cite{Gmiden2016Detection} the algorithm is only presented in a theoretical way, hence there is no test of the algorithm against any attack scenario, nor its detection performance are compared with previous works.
The computational overhead of the detection algorithm \Gmiden is equal to $\mathcal{O}(1)$ for each received CAN message.

\subsection{Song16}
In the work presented in~\cite{Song2016Intervals}, the authors design a detection algorithm based on the inter-arrival times of CAN messages. In particular, \Song evaluates the time difference $\Delta_{t}$ of messages with the same CAN ID and uses the time difference $\Delta_{t}$ for the detection of two attack scenarios.
The first attack scenario considered in \Song is the injection attack, in which messages are injected on the CAN bus randomly. To detect this attack scenario, the algorithm compares the time difference $\Delta_{t}$ with the cycle time $ct$ and raises an anomaly if $\Delta_{t}$ is lower than half the expected cycle time $ct$. We remark that this detection algorithm appears to be exactly the same as the one proposed in \Gmiden.
The second attack scenario is a Denial-of-Service attack, in which a message with a fixed value of the ID field is injected with a high frequency. For the detection of this attack scenario, the algorithm increments the value of a counter every time the $\Delta_{t}$ is lower than $0.2$ milliseconds, and raises an anomaly if the counter value is higher than a given threshold.
For the test of the algorithm, they used data gathered from a real vehicle and simulated the attacks by injecting messages for a random time window ranging from $5$ to $10$ seconds. In the injection attack scenario, they injected a message with twice, five, and ten times the original frequency, while in the DoS attack scenario the injection is fixed at $2000$ messages per second, testing threshold values of $1, 2, 3$, and $5$.
The detection results are evaluated by means of detection accuracy, but no comparison with previous work is presented.
The computational overhead of the detection algorithm \Song is equal to $\mathcal{O}(1)$ for each received CAN message.

\subsection{Moore17}
\label{ss:related_moore}
In the work presented in~\cite{Moore2017Frequency}, the authors describe a frequency-based anomaly detection algorithm.
The algorithm uses the time difference $\Delta_{t}$ between consecutive messages with the same ID value for its detection purposes. In particular, \Moore uses the sequence of $\Delta_{t}$ of each message ID for the identification of the \textit{maximum observed error} $m$, which is the maximum absolute difference between the expected cycle time $ct$ and the observed $\Delta_{t}$.
Upon reception of a message, the algorithm compares the $\Delta_{t}$ from the previous message with a threshold value defined as $ct \cdot 0.15 + m$. If three consecutive values of $\Delta_{t}$ are found outside the defined threshold, then an anomaly is raised.
The algorithm is tested against both injection and Denial-of-Service attacks, and presented the results by means of True Positive Rate (TPR), False Positive Rate (FPR), and False Negative Rate (FNR). No comparison with previous work is described.
The computational overhead of the detection algorithm \Moore is equal to $\mathcal{O}(1)$ for each received CAN message.

\subsection{Stabili19}
In the work presented in~\cite{Stabili2019Missing}, the authors designed an anomaly detection algorithm for the detection of missing messages from CAN communications. In particular, the algorithm presented in \Stabili evaluates the cycle-time ($ct$) of each CAN ID to build its detection model.
In the detection phase, the cycle time is used in conjunction with a configuration parameter $k$ (defined in the validation process for each ID) to detect missing messages from the CAN bus. A message with a particular ID is considered missing if it is not seen on the CAN bus for at least $ct \times k$ milliseconds.

This algorithm is tested against two similar attack scenarios, the \textit{ECU shutdown} attack (in which messages with a particular ID are removed from the CAN for a period of time that ranges from $10$ to $120$ milliseconds)  and the \textit{ECU inhibition} attack (in which all messages with a particular ID are removed from the CAN bus). The detection performance are presented by means of the $F$-measure and compared with other detection algorithms, despite only \Moore is used for the comparison with time-based anomaly detection algorithms.
The computational overhead of the detection algorithm \Stabili is equal to $\mathcal{O}(n)$ for each received CAN message, where $n$ is the number of message ID found in the monitored CAN section.

\subsection{Olufowobi20}
\label{ss:olufowobi20}
In~\cite{Olufowobi2020SAIDuCANT} the authors presented SAIDu-CANT, a specification-based intrusion detection system (IDS) using anomaly-based supervised learning.
The detection model is learned in the first phase of the algorithm, and requires a clean CAN data trace for learning different parameters that will be later used in the detection phase. The learned parameters are the minimum and maximum inter-arrival time of each CAN message ID $f_{i,min}, f_{i,max}$ , the estimated message period $\tilde{P_i}=f_{i,min}$ and the release jitter $J_i=f_{i,max}-f_{i,min}$.

In the detection phase the algorithm monitors the arrival time of each CAN IDs and compares it against the detection model. If a message is found outside the acceptable interval defined by the specification of the detection model, than it is labeled as anomalous.

The algorithm is tested against injection attacks on two different datasets. One dataset is composed by CAN traces gathered from two different Sedan vehicles, while the second dataset contains data from a single vehicle. Of these datasets, only the latter is publicly available\footnote{\url{https://sites.google.com/a/hksecurity.net/ocslab/Datasets/CAN-intrusion-dataset}}. 

The detection results are presented by means of True Negative (TN), True Positive (TP), False Positive (FP), False Negative (FN), accuracy, recall, precision, and F1 score. No comparison with previous work is presented, despite the authors showcased the detection performance of the algorithm against two other detection mechanisms directly described in the original work.
The computational overhead of the detection algorithm \Olufowobi is equal to $\mathcal{O}(1)$ for each received CAN message. 

%% file: sections/dataset.tex
\section{Dataset description}
\label{s:dataset}

This section describes the dataset used for the training and the test of the detection algorithms. We used two different datasets for testing and training the detection algorithms. The first dataset is gathered from the CAN bus of an unmodified, licensed $2016$ Volvo $V40$ Kinetic, and is called the \textit{Ventus} dataset. The second dataset is the OTIDS~\cite{Lee2017OTIDS} dataset, which is gathered from a KIA Soul.  

\subsection{Ventus dataset}
\label{ss:ventus_dataset}
The CAN data is recorded by physically connecting a laptop to the On-Board Diagnostic (OBD-II) port with a PCAN-USB adapter by Peak System~\cite{PCAN} and a D-Sub to OBD-II cable. The high-speed CAN bus segment exposed on the OBD-II port of the vehicle contains data related to the \textit{powertrain}, hence it is possible to access to CAN data frames exchanged by the ECUs to control the dynamic of the vehicle.
The Ventus dataset is composed by the \textit{clean} and the \textit{infected} sections. The former one is used for training and validating the detection algorithms, while the latter is used for the performance evaluation of the detection algorithms. For the generation of the \textit{infected} dataset, we build a threat model based on the attack scenarios considered by the analyzed algorithms. The final dataset is publicly available at~\cite{Pollicino2021TCPS}.

\subsubsection{Clean dataset}
\label{sss:clean_dataset}
The clean dataset is composed by $7$ different CAN traces, including more than $8$ million CAN messages corresponding to approximately $90$ minutes of CAN traffic. The CAN traces are gathered in different driving sessions performed on different road types (urban, suburban, and highway), traffic conditions, weather conditions, and geographical areas (plain, hill, and mountain). The CAN traces include ID, DLC, and payloads of each CAN data frame associated with its relative timestamp.
The clean dataset includes $51$ different message IDs, each one characterized by its own cycle time. The cycle time of each message ID is available to car makers and their suppliers in the \textit{DataBase for CAN (DBC)} file, which is used to describe details of CAN communications. However, this file is kept confidential, and is not publicly available. Since the cycle time of each message might influence the detection outcome, we need to estimate the cycle times of the messages in the clean dataset for the definition of multiple attack scenarios, each one targeting a message with a different cycle time. The cycle time of each message is evaluated as the mean value of the inter-arrival times between two consecutive messages with the same ID, and is rounded to the nearest millisecond. 

\begin{figure}[hptb]
    \centering
    \includegraphics[width=0.98\columnwidth]{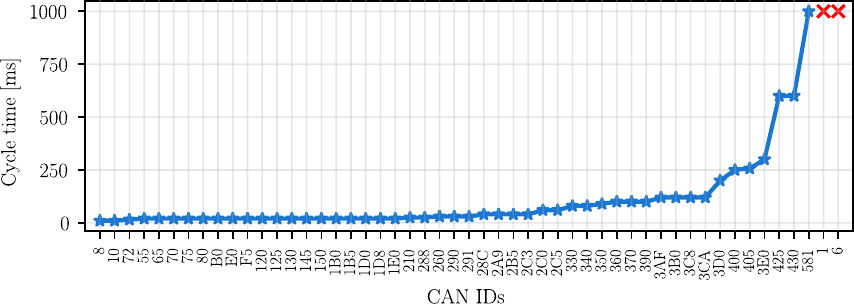}
    \caption{Cycle time evaluated on the Ventus clean dataset}
\label{fig:clean_distribution}
\end{figure}

Figure~\ref{fig:clean_distribution} shows on the $y$-axis the cycle time $ct$ (expressed in milliseconds) evaluated for each message ID (depicted on the $x$-axis) on the clean dataset. As shown in Figure~\ref{fig:clean_distribution}, the clean dataset is composed by $49$ messages exposing a cyclic behavior, while $2$ messages are identified as non-cyclic. The $49$ cyclic messages can be grouped in $17$ different cycle time classes, ranging from a minimum of $10$ milliseconds (IDs $0x8$ and $0x10$) to a maximum of $1$ second (ID $0x581$). We remark that these results are achieved on the clean dataset at our disposal. We empirically verified that the cycle time evaluated on a single trace of the clean dataset present extremely low variance compared with the cycle time evaluated on the other traces, hence we are confident that the results presented in Figure~\ref{fig:clean_distribution} are representative of the real cycle time of the messages gathered from the same vehicle model.

\subsubsection{Infected dataset}
\label{sss:infected_dataset}

The threat model used for the generation of the infected dataset is built on the attack scenarios described in the related work~\cite{Otsuka2014CANSecurity,Taylor2015Frequency,Gmiden2016Detection,Song2016Intervals,Moore2017Frequency,Stabili2019Missing}. The threat model is composed by two different attack scenarios, the \textit{message injection} and the \textit{message removal} attacks. The attack traces are generated in a laboratory environment for safety reasons. The laboratory setup is composed by a laptop computer, a Raspberry Pi $4$ board, and a CANPico~\cite{CANPico} device. The CAN bus is implemented through a breadboard. The laptop is connected via a PEAK CAN-USB device used to record the content of the CAN bus, the Raspberry is connected to the CAN bus via a CAN shield and is used to replay the normal traces gathered directly from our vehicle. The CANPico device has an integrated CAN transceiver and is connected directly to the CAN bus to generate the attacks. Since all transmitting devices are connected to the CAN bus via a CAN transceiver, re-transmissions, delays, arbitration and,
in general, all the low level details that might have been
affected by simulation artifacts are handled directly by the transceivers.
Since the algorithms considered in this work are based on the analysis of the timings of the CAN messages, each attack scenario is replicated by targeting $10$ different messages, each characterized by a different inter-arrival time. The list of the selected message IDs and their cycle-time are presented in Table~\ref{tab:id_details}.

\begin{table}[hptb]
    \centering
    \resizebox{0.98\columnwidth}{!}{%
    \begin{tabular}{@{}c|rrrrrrrrrr@{}}
    \toprule
    \textbf{ID}         & 0x10  & 0x120 & 0x290 & 0x2B5 & 0x2C5 & 0x330 & 0x350  & 0x3D0  & 0x3E0  & 0x581 \\ \midrule
    \textbf{Cycle Time} & 10 ms & 20 ms & 30 ms & 40 ms & 60 ms & 80 ms & 100 ms & 200 ms & 300 ms & 1 s   \\ \bottomrule
    \end{tabular}%
    }
    \caption{Details of the IDs selected for the simulation of the attack scenarios}
    \label{tab:id_details}
\end{table}

\textbf{Message injection}
The message injection attack scenario is used to inject messages on the CAN bus to subvert its normal behavior, by exploiting modern drive-by-wire capabilities such as automatic emergency braking, lane assist, park assist, adaptive cruise control, automatic transmission, and other similar features. 
The message injection scenario comprises different attacks, such as the \textit{replay attack}, in which a message already seen in the CAN communication is injected at a later time on the network, the \textit{fuzzing attack}, in which messages with altered field values are injected to study the consequences on the system, and the \textit{denial-of-service}, in which messages with high priority are injected with a very high frequency to prevent the delivery of normal CAN data frames.
For the aim of our analysis, we do not focus on both \textit{fuzzing} and \textit{denial-of-service} for the following reasons:
\begin{itemize}
    \item in case of a \textit{fuzzing attack} to the ID field of the CAN data frame, the detection algorithms would fail to detect any ongoing attack since it is not possible to build the detection model for a never observed message;
    \item the \textit{fuzzing attack} on the data field of the CAN data frame (with a legit message ID) can be considered as a \textit{replay attack} on the same message;
    \item the \textit{denial-of-service} attack scenario can be analyzed by considering a \textit{replay attack} with a high frequency of injection.
\end{itemize}
Hence, for the aforementioned reasons, we only focus on the \textit{replay attack} injection scenario. 
The replay attack is conducted by injecting on the CAN communication a target message (usually selected after an initial phase of reverse engineering of the values encoded in the payload) with a particular injection frequency. 

The final \textbf{injection attack} scenario considered in the experimental evaluation is composed by $50$ different attack instances, in which each of the $10$ selected message IDs is injected with a frequency of $1, 10, 25, 50$, and $100$ messages each second. The injected messages are equally distributed inside the $1$ second time window. Each attack scenario is simulated on all the traces of the clean dataset, for a total of $350$ CAN traces.

\textbf{Removal attack}
The removal attack scenario is used to remove messages from the CAN bus to subvert its normal behavior, preventing the ECUs from receiving data required for their functioning.
The removal attack is composed by two different attack scenarios, as already presented in~\cite{Stabili2019Missing}: the \textit{ECU shutdown} and the \textit{ECU inhibition} attacks. However, since the only difference between the two attacks is their duration, which does not impact on the overall performance of the detection algorithms, in this work we only focus on the \textit{ECU inhibition} scenario, in which target messages are completely removed from the CAN bus. 

The final \textbf{removal attack} scenario considered in the experimental evaluation is composed by $10$ different attack instances, in which each of the $10$ selected message IDs is completely removed from the CAN communication. As for the previous attack scenario, this attack is also simulated on all the traces composing the clean dataset, for a total of $70$ CAN traces.

\subsection{OTIDS dataset}
\label{ss:otids_dataset}
The OTIDS dataset is gathered from an unmodified, licensed KIA Soul vehicle~\cite{Lee2017OTIDS}. The OTIDS dataset is constructed by logging CAN traffic via the OBD-II port from a real vehicle while performing message injection attacks.
The OTDIS dataset is composed by $4$ different traces, one of them representing the clean dataset (\textit{``Attack free state''}) used for training of the algorithm, while the other traces represent $3$ different injection scearios. We also remark that all the traces composing the OTIDS dataset contain both data frame and remote frame messages. Since remote frames are not used in the algorithms tested in this paper and might change the outcome of the training process, we pre-processed the traces of the OTIDS dataset to leave only CAN data frames.

\subsubsection{Clean dataset}
\label{sss:otids_clean}
\begin{figure}[hptb]
    \centering
    \includegraphics[width=0.98\columnwidth]{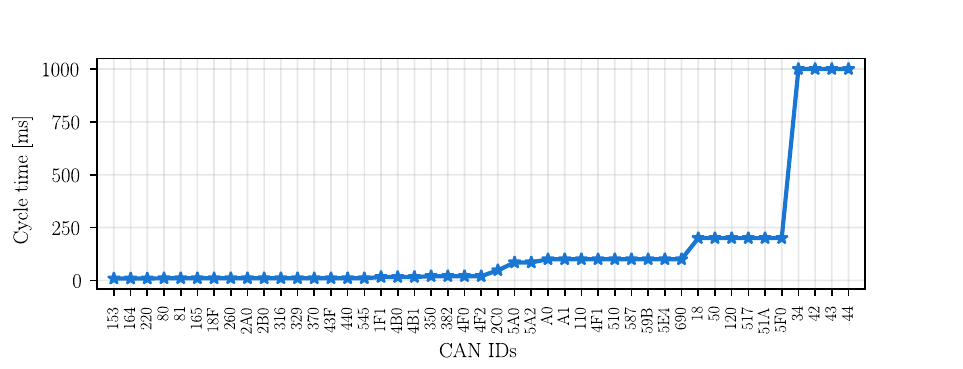}
    \caption{Cycle time evaluated on the OTIDS clean dataset}
\label{fig:otids_timestamps}
\end{figure}

The clean part of the OTIDS dataset is composed by a single trace, with a little more than $1.4$ million CAN messages corresponding to approximately $10$ minutes of CAN traffic. The clean CAN trace include the ID, DLC, and payloads of each CAN data frame associate with its relative timestamp. The clean part of the OTIDS dataset includes $45$ different message IDs, each one characterized by its own cycle time. As for the Ventus dataset, we extracted the cycle times of each message from the clean trace (with the same process already described in~\ref{sss:clean_dataset}), and the results of the analysis is shown in Figure~\ref{fig:otids_timestamps}. 

From the results shown in Figure~\ref{fig:otids_timestamps}, all the $45$ IDs found in the OTIDS dataset exhibit a cyclic behavior, with a minimum of $9$ milliseconds (IDs $0x153$, $0x164$, and $0x220$), to a maximum of $1$ second (IDs $0x34, 0x42, 0x43$, and $0x44$). In the OTIDS dataset we found $9$ different cycle time classes.

\subsubsection{Infected dataset}
\label{sss:otids_infected} 
The OTIDS dataset assumes a threat model extremely different from the one considered in the Ventus dataset. We remark that the OTIDS infected dataset is composed by three different types of Injection attacks:
\begin{itemize}
    \item \textbf{Fuzzy:} in the fuzzy attack scenario, messages of spoofed random CAN ID and data are injected in the CAN communication;
    \item \textbf{Denial-of-Service:} in the DoS attack scenario, messages with a CAN ID set to $0x0$ are injected in the CAN communication with a high frequency;
    \item \textbf{Impersonation:} in the impersonation attack scenario, valid messages with ID $0x164$ are removed from the network and the attacking node injects messages with the same ID. 
\end{itemize}

In case of fuzzy and impersonation attacks, the injection starts after $250$ seconds of normal CAN communication, while in case of the DoS attack scenario the attack starts at the beginning of the trace.

%% file: sections/implementations.tex
\section{Implementations of the detection algorithms}
\label{s:implementations}

We observe that none of the related work considered in this paper is distributed together with a reference implementation. Moreover, several papers neglect many relevant details that are actually required to implement the proposed detection algorithm. In this section, we describe the additional assumptions used for the implementation of the detection algorithms. All assumptions are motivated and based on the maximization of the detection capabilities of the algorithms. We remark that all our reference implementation are publicly available at~\cite{Pollicino2021TCPS}, thus allowing researchers to easily replicate our experiments and benchmark novel time-based detection algorithms with respect to the state-of-the-art.

\subsection{Implementation of Otsuka14}
\label{ss:implementation_otsuka}

\begin{figure}[tbh]
    \centering
    \includegraphics[width=0.98\columnwidth]{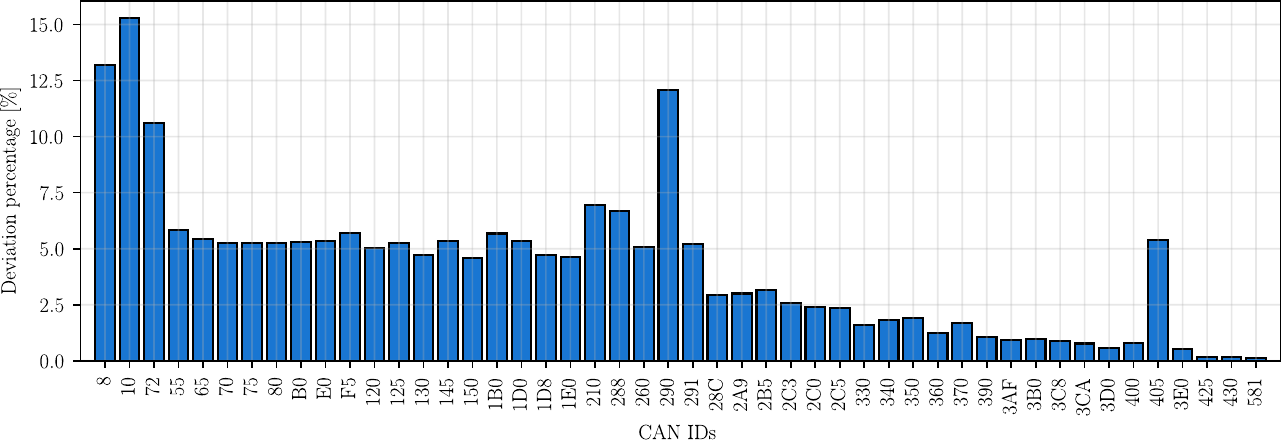}
    \caption{Deviation percentage on the Ventus dataset for each CAN ID}
    \label{fig:ventus_otsuka_deviation}
\end{figure}

\begin{figure}[tbh]
    \centering
    \includegraphics[width=0.98\columnwidth]{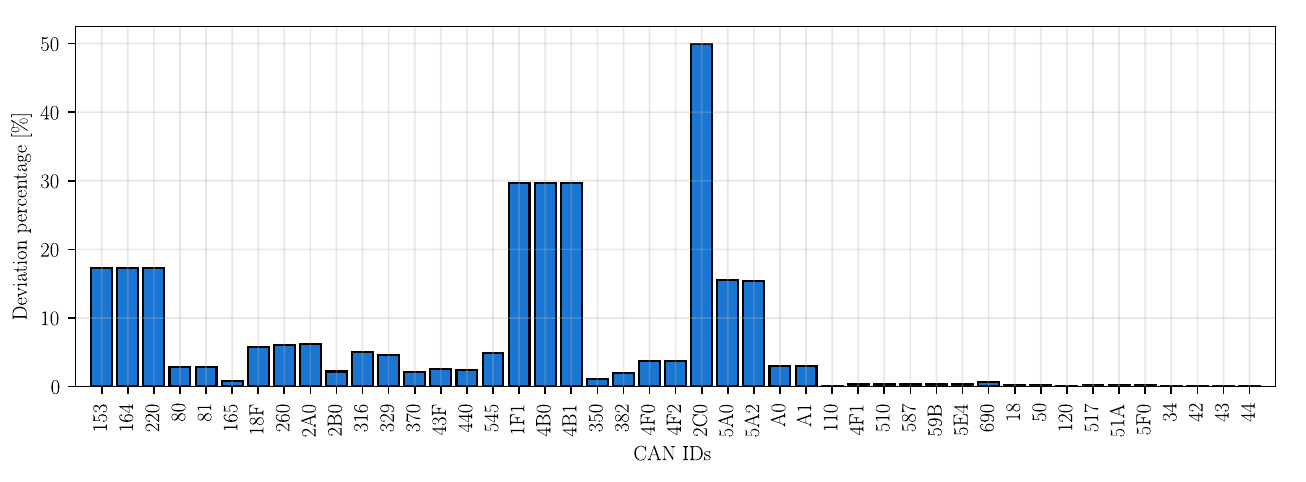}
    \caption{Deviation percentage on the OTIDS dataset for each CAN ID}
    \label{fig:otids_otsuka_deviation}
\end{figure}

For the implementation of the \Otsuka detection algorithm we used the estimated cycle times (see Figures~\ref{fig:clean_distribution} and~\ref{fig:otids_timestamps}) as the \textit{mean reception cycle} used in the detection process. We remark that the original work only considers $2$ IDs, of which only $1$ is cyclic. The cyclic ID has a maximum deviation with respect to the mean cycle time of $2\%$, while the other ID has a maximum deviation of $~30\%$. Based on these results, authors set a fixed detection threshold of $\delta = 5\%$ from the expected cycle time. Hence any message received outside the valid time range of [$ct^{ID}-5\%$, $ct^{ID}+5\%$] is considered anomalous. The value of the threshold $\delta$ is defined as the threshold value that minimizes the number of false positives in the validation process.
We replicated the same experimental analysis on both datasets, and the results are presented in Figure~\ref{fig:ventus_otsuka_deviation} and~\ref{fig:otids_otsuka_deviation} for the Ventus and OTIDS dataset, respectively. The two figures shows the deviation from the evaluated cycle time for each message ID. The IDs of the messages are sorted by their cycle time in ascending order (left to right). The results depicted in both Figures~\ref{fig:ventus_otsuka_deviation} and~\ref{fig:otids_otsuka_deviation} show that messages found on the CAN bus more often exhibit a higher deviation from the expected cycle time.

The value of $\delta$ that minimizes the overall number of false positives on the Ventus dataset is experimentally evaluated to be $\delta = 4\%$. While testing the best value of $\delta$ to minimize the false positives on OTIDS however, we found out that by increasing the value of $\delta$ improves the output of the validation results. However, since higher values of $\delta$ introduce a larger time window in which messages are not considered anomalous, the value used in the experimental evaluation is fixed at $\delta = 25\%$. In the experimental evaluation presented in the next section we configured the \Otsuka algorithm with these values for its detection purposes. 

Moreover, the algorithm presented in \Otsuka is designed to discard the malicious CAN messages by using a waiting mechanism. With this mechanism, all the messages with the same CAN ID received before $ct + \delta$ are held and discarded as soon as another message with the same ID is received after $ct + \delta$. However, this mechanism is based on the assumption that at least the first received message for each ID is legit, and its detection performance are highly affected in case this assumption is violated. To this aim, our implementation of \Otsuka considers all the messages with the same ID received before $ct + \delta$ as a single case: in case at least one of the held messages is an injected frame then a single anomaly is raised, while if no one of the held messages is anomalous only one false positive is considered. This design choice minimizes the false positive rate of this algorithm in case of high frequency injection attacks, thus optimizing its detection results.

\subsection{Implementation of Taylor15}
\label{ss:implementation_taylor}
For the implementation of \Taylor we used the cycle times estimated on the clean datasets as the \textit{historical mean} required by the algorithm for the $t-$tests. For the experimental evaluation of the detection performance of \Taylor we follow the same assumptions described in~\cite{Taylor2015Frequency}. 
We remark that in \Taylor the messages with a cycle time below $50$ms are representative of more than $90\%$ of the CAN IDs, while in the Ventus and the OTIDS datasets these messages are only $61.22\%$ and $53.33\%$ respectively of the whole datasets ($30$ out of $49$ and $24$ out of $45$ cyclic messages). This difference does not impact the detection performance of the algorithm, however this method is applicable only to messages having a cycle time below $50$ms, hence in both datasets the \Taylor algorithm has limited applicability compared to the original paper.
Moreover, in the original work where \Taylor is presented there are some missing details that prevent it from being directly implemented, hence some additional assumptions are made also for this algorithm.

The first additional assumption is focused to the number of sequences $\mathcal{S}_q$ used for the evaluation of the anomaly score $A(\mathcal{S}_q)$. This is a crucial aspect for the evaluation of the detection performance of the algorithm, since it impacts directly on the detection performance. As an example, consider a scenario in which a value of $\mathcal{S}_q = 10$ is used, i.e. the anomaly score is evaluated every $10$ $t-$tests. As described in~\cite{Taylor2015Frequency}, each item of the sequence is the value of the highest $t-$test in a window of $1$ second. Hence, in the considered scenario, an anomaly score $A(\mathcal{S}_q)$ evaluated on a sequence of $10$ $t-$tests implies that the algorithm is able to detect a single anomaly within a $10$ seconds time window. This design choice would prevent a fair comparison between the detection performance of \Taylor and the other algorithms, that are able to flag many different anomaly in a similar time window.
We tested different values of $\mathcal{S}_q$ for the evaluation of the anomaly score, from a minimum of $\mathcal{S}_q = 2$ up to the whole series of t-tests. At the end of this analysis we discovered that the mean value of the $A(\mathcal{S}_q)$ is minimally affected by the number of sequences used for its evaluation. Hence, we chose the value of $\mathcal{S}_q = 2$ to allow the detection of anomalies in the smallest possible time window.

\begin{figure}[hptb]
    \centering
    \includegraphics[width=0.98\columnwidth]{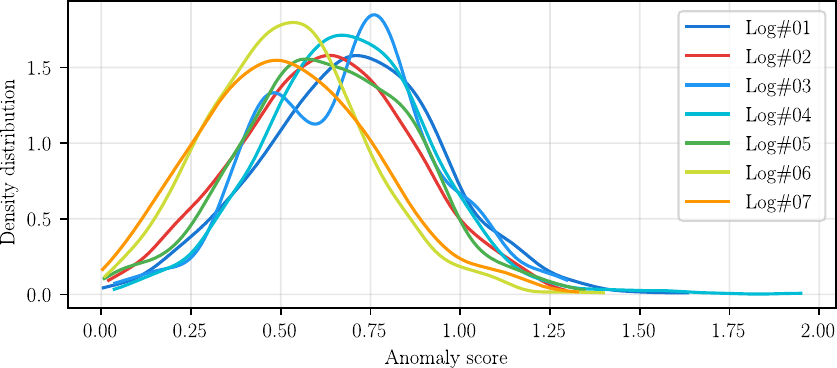}
    \caption{Distribution of the $A(\mathcal{S}_q)$ on the Ventus dataset}\label{fig:ventus_taylor_validation}
\end{figure}

\begin{figure}[hptb]
    \centering
    \includegraphics[width=0.98\columnwidth]{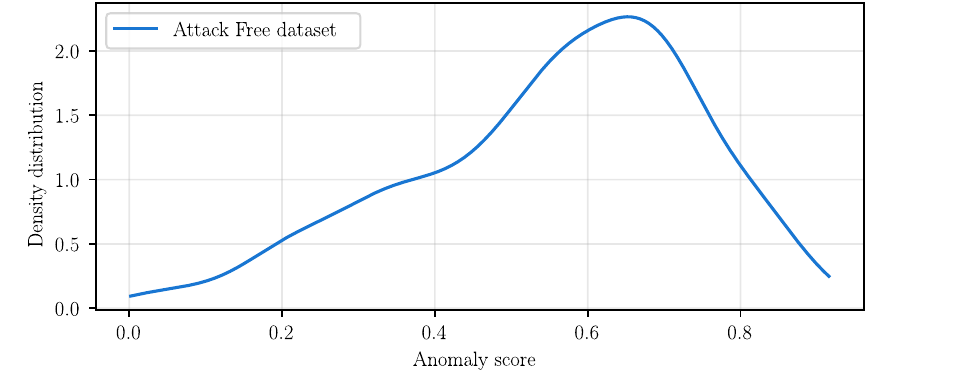}
    \caption{Distribution of the $A(\mathcal{S}_q)$ on the OTIDS dataset}\label{fig:otids_taylor_validation}
\end{figure}

The second additional assumption is focused on the definition of the detection threshold. In~\cite{Taylor2015Frequency} this aspect is not discussed since the presented evaluation is focused on the analysis of the ROC and AUC measures, which are threshold-independent. However, for the performance evaluation we need to define a threshold value used to distinguish between normal and anomalous time windows.
As a first step in the definition of the threshold value we analyzed the distribution of the anomaly scores on the clean datasets, and the results are shown in Figure~\ref{fig:ventus_taylor_validation} for the Ventus dataset and in Figure~\ref{fig:otids_taylor_validation} for the OTIDS dataset.

From the analysis of the distribution of the anomaly scores presented in Figures~\ref{fig:ventus_taylor_validation} and~\ref{fig:otids_taylor_validation} it is possible to design two different detection mechanisms for each dataset.
The first one is based on the analysis of the distribution range across the different traces, and defines a threshold value higher than the maximum anomaly score evaluated in the validation process. As an example, by considering the results achieved on the Ventus dataset depicted in Figure~\ref{fig:ventus_taylor_validation}, the anomaly scores are always below $2.0$, hence it is possible to use this value as the threshold for the classification of the anomalies. In the detection process, any $A(\mathcal{S}_q) \geq 2.0$ is considered as an anomaly and all values below the threshold as considered legit values.
The second detection mechanism is based on the similarity of the distributions with the distribution of the normal function, hence it is possible to define a threshold using the mean value and the standard deviation of the anomaly scores, as already proposed in~\cite{Chandola2012Discrete, Marchetti2016Entropy}. Anomaly detection methods based on the similarity with the normal distribution often define the normal threshold as [$\mu - 3 \times \sigma$, $\mu + 3 \times \sigma$], and consider any outside value as anomalous. However, we remark that only $99.68\%$ of the distribution is covered with a value of $k = 3$, hence an anomaly detector based on this detection model introduces $0.32\%$ of false positives.
Since this last detection method introduces false positives that negatively impact the performance evaluation of this detector, our implementation of \Taylor uses a fixed detection threshold of $2.0$ for the Ventus dataset and of $1.0$ on the OTIDS dataset, to distinguish between valid and anomalous time windows. 

\subsection{Implementation of Cho16}
\label{ss:implementation_cho}
The \Cho detection algorithm uses the estimated clock skew for the detection of attacks to the CAN communications. The algorithm used for the clock skew estimation is described in the original work (Algorithm~$1$) and, despite the pseudo-code is well documented and described, there are a few missing information that require additional assumptions. 
The most critical assumptions are related to the initialization of different parameters used for the evaluation of the identification error. Since the identification error is used for the detection task, it is critical to initialize these values correctly. The three most impacting parameters are the size of the window $N$, the initial value of the parameter $P$ used in the \texttt{SkewUpdate} procedure, and the number of standard deviations $\kappa$. While the size of the window $N$ only impacts the detection delay of the algorithm, the initial value of the parameter $P$ heavily impacts on the computation of the identification error, thus impacting the detection capabilities of the algorithm. In an example described in the original work $N$ and $\kappa$ are initialized to the values of $20$ and $5$, respectively. However, the value used to initialize $P$ is not disclosed. We tested different combinations of $N$, $P$, and $\kappa$ on the clean datasets, aiming to minimize false positives false positives. The final values evaluated on the Ventus dataset are $N = 10$, $P = 0.05$, and $\kappa=0.1$, while on the OTIDS dataset the lowest number of false positives is given by the combination of $N = 5$, $P = 0.001$, and $\kappa=2.5$. We remark that these values are identified after a long process of testing on the clean dataset, since there is no description on how to identify the best parameters in the original work.  
Moreover, since the threat model considered in~\Cho assumes that the attack starts after $420$ seconds of normal CAN communication, and the attacks on the Ventus dataset are generated from the beginning, we modified our implementation to allow \Cho to evaluate the clock skew and identification errors on the clean base trace used for  attack generation. This allows \Cho to learn the clock skew and identification errors on legit values, which are then used as reference against the attack scenarios.

\subsection{Implementation of Gmiden16}
\label{ss:implementation_gmiden}
The description of the \Gmiden detection algorithm presented in its original work is complete of all the details required for its functioning, hence we were able to produce a reference implementation that comply with the original design without the need for additional assumptions or design choices. In our version, we used the cycle time evaluated on the clean dataset as the reference inter-arrival time required by the algorithm.

\subsection{Implementation of Song16}
\label{ss:implementation_song}  
The \Song detection algorithm presented in~\cite{Song2016Intervals} is based on the assumption that CAN messages exhibit a fixed inter-arrival time. However, the experimental analysis presented in~\ref{ss:implementation_otsuka} demonstrate that this assumption does not hold on our dataset, where more frequent messages have a higher deviation from the mean value. We remark that this phenomena is well known in real CAN buses implemented in modern vehicles, also other real datasets exhibit the same behavior, due to possible delays and re-transmissions in CAN bus segments that have a relatively high usage (about $50\%$ or higher). This is also acknowledge by producers of automotive ECUs, that consider deviations up to $15\%$ of the reference cycle time to be within the normal working parameters. 
While these assumptions are included in the original paper, it appears that they are not necessary for the proposed detection algorithms.

We recall that \Song includes two different algorithms targeting message injection and denial of service, respectively. The algorithm for detecting message injection appears to be identical to \Gmiden, hence we reuse the same implementation.
The algorithm for detecting DoS attacks is clearly explained in the original paper, hence we produced a reference implementation that fully complies with the description provided by the authors.

\subsection{Implementation of Moore17}
\label{ss:implementation_moore}
In the original work presenting \Moore, the authors assumed that the first $15$ seconds of each trace is unaltered, and they used the first $5$ seconds for the definition of the detection model. However, in the infected traces composing the Ventus dataset the attacks are generated starting at the beginning of each trace, as described in Section~\ref{sss:infected_dataset}. 
Since the Ventus dataset contains a clean dataset composed by more than $90$ minutes of clean CAN traffic, we used the first $5$ seconds of the traces composing the clean dataset for training \Moore. However, in the OTIDS dataset the attacks are generated after $250$ seconds of normal traffic, hence we trained the detection model on the first $5$ seconds of the infected trace.

\begin{figure}[hptb]
    \centering
    \includegraphics[width=0.98\columnwidth]{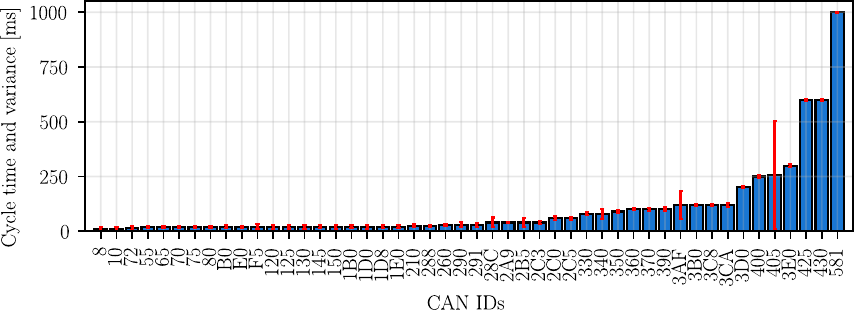}
    \caption{Maximum distance from average reception cycle [ms] on the Ventus dataset}\label{fig:ventus_moore_max}
\end{figure}

\begin{figure}[hptb]
    \centering
    \includegraphics[width=0.98\columnwidth]{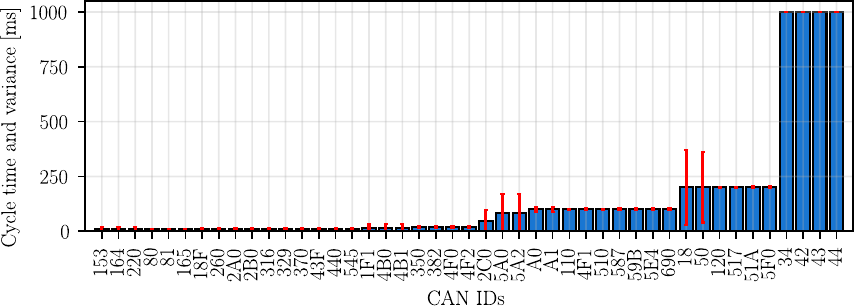}
    \caption{Maximum distance from average reception cycle [ms] on the OTIDS dataset}\label{fig:otids_moore_max}
\end{figure}

Following the analysis presented in~\cite{Moore2017Frequency}, we also compared the maximum variance of the cycle time of each message ID with respect to the expected cycle time $ct$ on the first $5$ seconds of the traces composing the datasets. Results are presented in Figure~\ref{fig:ventus_moore_max} and~\ref{fig:otids_moore_max} for the Ventus and OTIDS dataset, respectively.

From this analysis it is possible to notice that all the IDs exhibit a small variance from the expected cycle time in the first $5$ seconds of the traces of the clean section of the datasets, with the exception of ID $0x405$ for the Ventus dataset and of IDs $0x18$ and $0x50$ for the OTIDS dataset.
We remark that these results are comparable to the ones presented in~\cite{Moore2017Frequency} on the dataset at their disposal. 

We recall that \Moore raises an alert if the time between consecutive messages with the same ID $\Delta_{t}$ is lower or grater than a threshold (see Section~\ref{ss:implementation_moore} for additional details).   
To prevent false positives, \Moore raises an anomaly only if three consecutive alerts are generated for the same ID. This design choice prevents a direct comparison of false positive and false negative rates with the other algorithms.

To allow performance comparisons without penalizing \Moore, given an anomaly we classify all the injected messages that generated one of the three consecutive alerts as a true positive. If a legit message generated one of the alerts required for issuing an anomaly, we do not count that as a false positive. On the other hand, if an anomaly has been raised after three alerts all generated by legit messages, we consider that anomaly as a single false positive. This solution increases the overall detection performance of \Moore and makes it comparable to other algorithms in terms of $\mathcal{F}-$measure.


\subsection{Implementation of Stabili19}
\label{ss:implementation_stabili}
\begin{figure}[hptb]
    \centering
    \includegraphics[width=0.98\columnwidth]{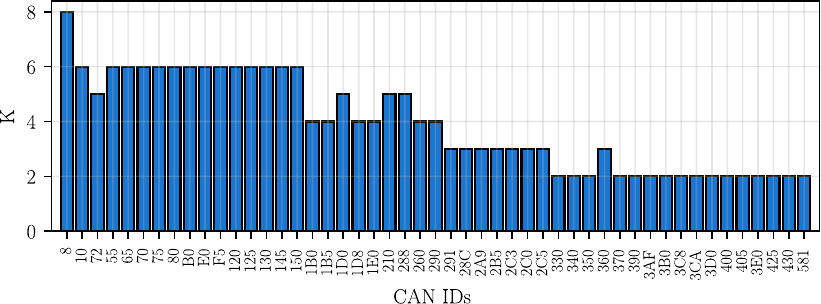}
    \caption{Comparison of the value of the configuration parameter $k$ with the $ct$ of each message ID on the Ventus dataset}\label{fig:ventus_stabili_k}
\end{figure}

\begin{figure}[hptb]
    \centering
    \includegraphics[width=0.98\columnwidth]{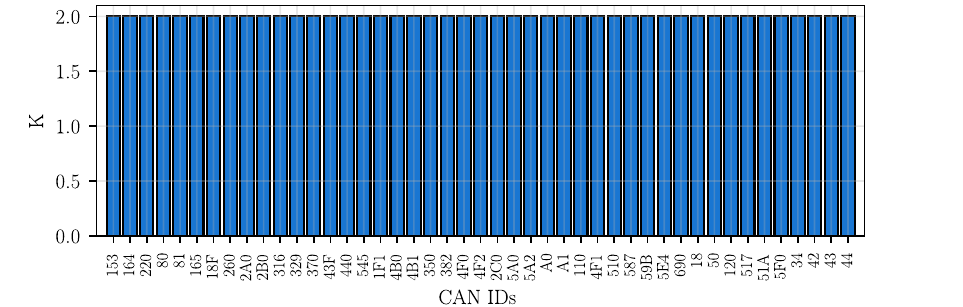}
    \caption{Comparison of the value of the configuration parameter $k$ with the $ct$ of each message ID on the OTIDS dataset}\label{fig:otids_stabili_k}
\end{figure}

The \Stabili detection algorithm is based on the assumption that it is possible to create a detection model that raises $0$ false positives in the validation phase. To this aim, \Stabili uses a parameter $k$ for the definition of the \textit{valid waiting time} for each different ID. 
We replicated the tuning of the parameter $k$ following the description provided in~\cite{Stabili2019Missing} over both datasets, and the results of this analysis are presented in Figure~\ref{fig:ventus_stabili_k} and~\ref{fig:otids_stabili_k} for the Ventus and OTIDS datasets, respectively.

At the end of this training process on the Ventus dataset, the minimum value of the parameter $k$ that does not generate false positives on the clean dataset is equal to $2$ (for $16$ message IDs), while the maximum value is $8$ (for a single message ID). On the OTIDS dataset however, all the IDs do not generate any false positive by using a value of $k = 2$.

Moreover, focusing on the Ventus dataset and comparing the results presented in Figure~\ref{fig:ventus_stabili_k} with the ones presented in Figure~\ref{fig:ventus_otsuka_deviation} it is possible to notice that messages with lower cycle time values exhibit a higher deviation from the expected cycle time and require a higher value of the parameter $k$ to achieve $0$ false positives. This interesting result is explained by considering that messages with lower cycle time (i.e. appearing on the CAN bus more frequently than the others) are more prone to delays and transmission errors over a real CAN bus that has a load comparable or higher to $50\%$, hence they require higher tolerance by any detection mechanism based on the analysis of the timings. 



\subsection{Implementation of Olufowobi20}
\label{ss:implementation_olufowobi}
The \Olufowobi detection algorithm is composed by the training and the detection phase, which are presented in Algorithm~$1$ and~$2$ of the original paper, respectively.


The algorithm is based on the assumption that the real model and its parameters are unknown and unavailable, since these information (specifically the precise message periods) is generally not disclosed by manufacturers. 
For this reason \Olufowobi considers a detection model and based on parameters derived from observations of the CAN communications. Hence, the training phase of \Olufowobi is focused on the definition of the parameters required for the detection process: the estimated message period $\tilde{P_i}=f_{i,min}$ and the release jitter $J_i=f_{i,max}-f_{i,min}$. 

For the estimation of these parameters the transmission time $C_i$ is required, which is defined as the time required to transmit a message on the CAN bus. However, we remark that it is not possible to identify the value of $C_i$ by observing normal CAN communications but should be learned by measuring the transmission time on the ECU transmitting a CAN message.  
Since it is not possible to identify the value of $C_i$ from the clean dataset, we used a worst-case estimation of this parameter in the training process. For the worst-case estimation of $C_i$ we used the clean dataset to reconstruct the sequence of bits composing the CAN data frame to calculate the size of the message $size(m_i)$. Than the value of $C_i$ is defined as $C_i = max(size(m_i)) / bitrate$, where $max(size(m_i))$ is the maximum reconstructed bit size of CAN messages with the same ID $i$ and $bitrate$ is the nominal bitrate of the CAN. In our experiments, the nominal bitrate of both datasets is $500kbps$. 

Another issue raised during the implementation of \Olufowobi is related to the availability of the value of the precise message period $P_i$. We remark that $P_i$ (as described in the original work) belongs to the list of parameters not available for the definition of the detection model. However, in the description of the detection process (Algorithm~$2$ of the original work) the precise message period $P_i$ is used as one of the parameters of the algorithm. Since the precise message period $P_i$ is different from the estimated message period $\tilde{P_i}$ (which is computed in the training phase), in our implementation we defined $P_i$ as the cycle time $ct$.
 
Moreover, another issue raised in the implementation of the detection algorithm we discover that the variable $k$ (one of the inputs of the detection function) is modified (in Algorithm $2$, row $7$) but the modified value is used. This variable is used for the evaluation of the arrival time window of the next message, and is crucial for the detection task. In our implementation of \Olufowobi we modified the detection algorithm by returning the value of $k$ in case is modified. 

Despite the aforementioned issues raised during the implementation of \Olufowobi, in its experimental evaluation we discovered that the latter modification introduced a significant downside that afflicts the detection performance of \Olufowobi. In particular, the novel issue is related to the modification of the value of $k$ in case a message is marked as an anomaly. In this scenario, the algorithm will update the value of $k$ and, in case of injection attacks, all the following messages are identified as anomalies since the arrival time window used for the detection task is out of sync with the tested message, resulting in thousands of false positives. To overcome this issue, we introduced an ``update protection mechanism'' on the value of $k$ that triggers when a message is considered anomalous, reverting the value of $k$ to its previous value.

%% file: sections/experimental_evaluation.tex
\section{Experimental evaluation}
\label{s:experimental_evaluation}

In this section, we present the experimental evaluation of the detection performance of the algorithms against the datasets (and their respective threat models) presented in Section~\ref{s:dataset}.
To compare the detection performance of the different algorithms against the described attack scenarios we consider the \fmeasure as the key performance indicator. The \fmeasure is the harmonic mean of the precision and the recall of the detection performance. This indicator is commonly used in comparing intrusion detection systems. It ranges from $0$ to $1$, where values close to $0$ denote the inability to detect any anomaly and values close to $1$ denote the ability of the algorithm to detect all the anomalies with low false positives. The perfect detection algorithm exhibiting $100\%$ precision and recall has \fmeasure equal to $1$.

\subsection{Detection results against the Ventus dataset}
\label{ss:detection_ventus}

\subsubsection{Message injection detection comparison}
\label{sss:injection_comparison}
\begin{figure*}[hptb]
    \centering
    \includegraphics[width=\textwidth]{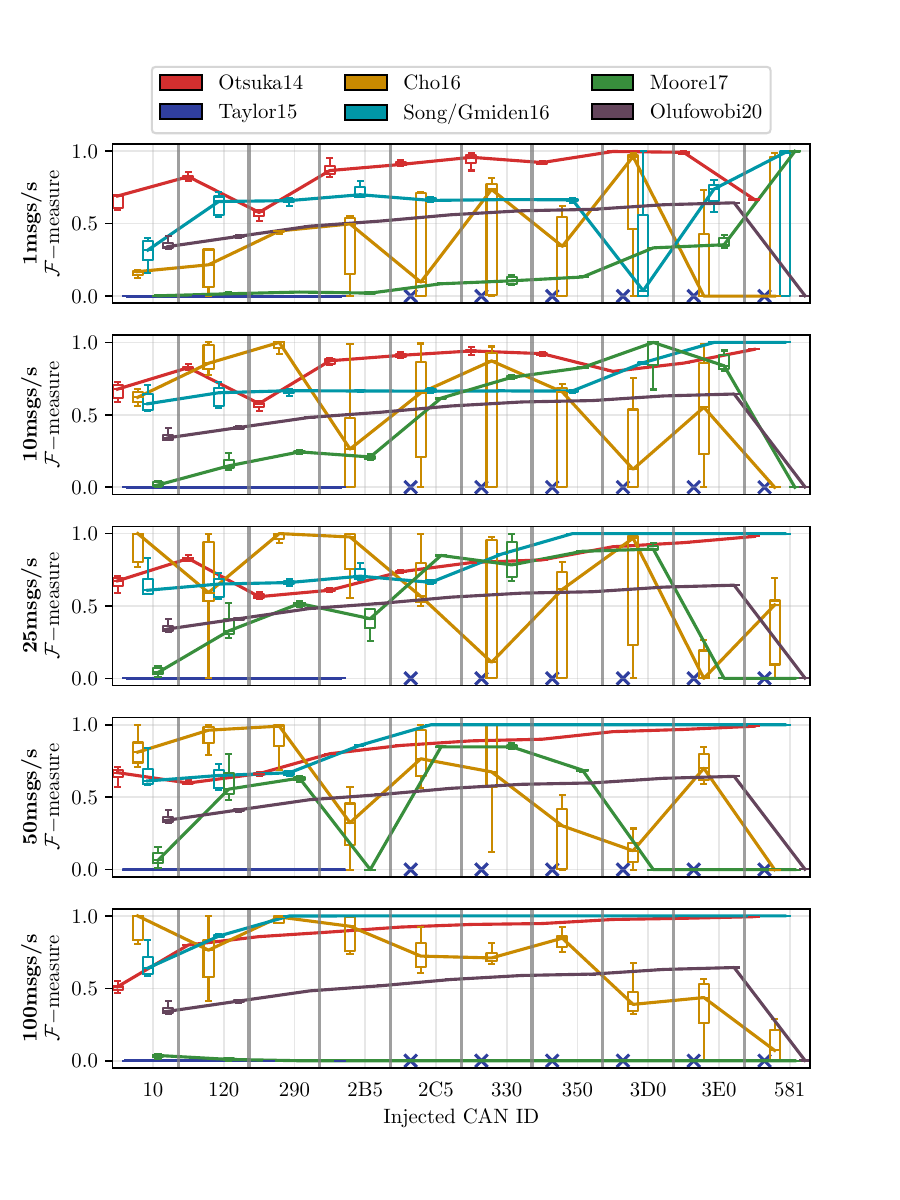}
    \caption{Experimental evaluation of the detection algorithms against the injection attack scenario of the Ventus dataset}\label{fig:injection_results}
\end{figure*}

The performance evaluation of the algorithms against the message injection attack is presented in Figure~\ref{fig:injection_results}. Figure~\ref{fig:injection_results} is composed by $5$ different subplots representing different attack scenarios. Top to bottom, these attack scenarios are the injection of $1, 10, 25, 50$, and $100$ messages each second.
The $y$-axis of each subplot of Figure~\ref{fig:injection_results} shows the \fmeasure evaluated with the detection algorithms, while the $x$-axis represents the injected message ID, ordered by ascending cycle time. The $10$ reference message IDs used in these experiments are the ones presented in~\ref{tab:id_details} of Section~\ref{sss:infected_dataset}, as they represent messages with different cycle times.

The detection results of each algorithm against a given combination of ID and injection frequency is represented as a boxplot. We used this graphical indicator since it allows to summarize in a concise representation the detection results achieved against the same attack (same ID and injection frequency) across different infected traces. For readability purposes, we only show the main components of the boxplot: the $10_{th}, 25_{th}, 50_{th}$ (median), $75_{th}$, and $90_{th}$ percentiles. 

As an example, the top subplot of Figure~\ref{fig:injection_results} refers to injection attacks performed with an injection frequency of $1$ message per second. This subplot is divided into $10$ ``columns'', each referring to the injection of a different message ID. The first of these columns refers to the injection of message ID $0x10$, and includes $6$ boxplots drawn in different colors and having a small horizontal offset to improve readability. The first boxplot (red) summarizes the detection performance of \Otsuka, the second (blue) of \Taylor, the third (dark orange) of \Cho, the fourth (cyan) of \Song and \Gmiden, the fifth (green) of \Moore, while the last one (purple) of \Olufowobi. We remark that \Stabili is not included in this set of experiments since it is designed to detect only missing messages, and it cannot be applied against injection attacks. To highlight the trend related to the detection performance depending on the message cycle time, we also draw solid lines connecting the median values of all the boxplots related to the same algorithm. 

We recall that \Taylor is designed to support only messages with a cycle time lower than $50$ms (see Section~\ref{ss:implementation_taylor} for additional details), hence this algorithm is only applicable to the first $4$ message IDs having the lower cycle times among the $10$ considered in our experiments. To visually represent the inapplicability of \Taylor to the last $6$ IDs we draw a blue $\times$ instead of the boxplot.
 
From the analysis of the results shown in Figure~\ref{fig:injection_results} it is possible to notice different trends by either focusing on the comparison of the detection performance against the injection of the same volume of messages with different IDs (``horizontal'' analysis) or by focusing on the difference between the performance against the increasing injection rate of the same message ID (``vertical'' analysis). In both cases however, we remark that our implementation of \Taylor is not able to detect any anomaly in the considered attack scenarios, since the injection of messages does not introduce a significant deviation from the normal anomaly score used by this algorithm. 

With the ``horizontal'' analysis it is possible to notice that by using \Otsuka and \Song (\Gmiden) the overall detection performance of the algorithms increases by simulating the attack on messages with higher cycle times, while the overall trend for \Cho is the opposite. By focusing on the results of \Otsuka and \Song (\Gmiden) algorithms with a ``vertical'' analysis it is possible to notice that \Otsuka achieves higher detection performance against the injection of a low volume of messages ($\leq 10$ messages per second), while for injection frequency of at least $25$ messages per second \Song (\Gmiden) converge to higher \fmeasure also against the injection of messages with low cycle times. Both algorithms are limited in their detection against the injection of the message with ID $0x10$ (i.e. the message with the lowest cycle time in our dataset). We remark also that \Song (\Gmiden) is able to achieve a \fmeasure of $1.0$ against different injection scenarios (the top $2, 4, 6$, and $8$ messages with the highest cycle times against the injection of $10, 25, 50$, and $100$ messages per second, respectively), while \Otsuka achieves a perfect \fmeasure only in a subset of these scenarios (the top $1$, and $3$ messages with the highest cycle times against the injection of $50$, and $100$ messages per second, respectively). 
By focusing on \Cho with a ``vertical'' analysis however it is possible to notice that the overall detection performance of the algorithm increases by increasing the injection frequency, despite having high variance across different tests. 

The detection results achieved by \Moore however are completely different and require a dedicated analysis. As presented in Figure~\ref{fig:injection_results}, by comparing the detection performance of \Moore against the different injection frequencies it is not possible to identify any trend with both ``horizontal'' and ``vertical'' analysis. 
We recall that \Moore uses a false-positive prevention system which raises an anomaly only after $3$ consecutive alerts, as described in Section~\ref{s:related}. 
As an example, consider the case of injection of a single message each second using the message with the lowest cycle time (ID $0x10$). In this scenario the detection performance of \Moore are extremely low since the injected message is right after or just before another valid message, thus resulting in the generation of up to $2$ alerts, that are not enough to cause the generation of an anomaly.
However, messages with higher cycle times have a smaller margin of error (comparing the experimental results for the definition of the parameter $m$ discussed in Section~\ref{ss:implementation_moore}), hence it is possible to detect anomalies more frequently.
One could expect that by increasing the injection frequency the overall detection performance should also increase, however the experimental evaluation demonstrates that for high enough injection frequencies \Moore starts to generate an increasing number of false negatives (i.e. missed anomalies). To better explain this counter-intuitive behavior, we refer to an example. Consider the scenario of the injection of the message $0x581$, which has a $ct = 1000$ms. For this message, the value of the parameter $m$ is $20$ms, hence the time required for the identification of an alarm for this message is approximately $152$ms (see Section~\ref{ss:related_moore} for additional details on how \Moore detects anomalies). Since in our dataset the injection is simulated by equally distributing the injected messages on the interested time window, by injecting messages with a frequency of $10$ messages per second we are injecting a message every $100$ milliseconds, thus the time required for the identification of an alert of $152$ms is higher than the cycle time between two consecutive messages, increasing the number of false negatives. This explains the two different trends that can be observed by comparing the ``vertical'' and ``horizontal'' analysis. By targeting messages with higher cycle time, the overall detection performance of \Moore increases. However, by increasing the injection frequency it is possible to increase the detection performance against the injection of messages with lower cycle times, despite there is a huge increment of false negatives for messages whose time required for the identification of a single alert is higher than the time between two consecutive injected messages. 

Finally, by analyzing the detection performance of \Olufowobi ``horizontally'' it is possible to notice that the detection performance of the algorithm increases by increasing the cycle time of the injected message, in a trend similar to the one already observed for \Otsuka and \Song (\Gmiden). However, with the ``vertical'' analysis it is possible to notice that the detection performance are not influenced by increasing the frequency of the injection. We also remark that the detection performance of \Olufowobi against the injection of the message with the highest cycle time are always equal to $0$.

\subsubsection{Message removal detection comparison}
\label{sss:removal_comparison}
To compare the detection performance of the only two algorithms supporting the detection against the message removal attack we present the results by means of \textit{percentage of detected anomalies}, i.e. the number of alarms raised by the two detectors compared to the number of removed messages. We remark that it is possible to use this detection metric only in case the training of the detectors is based on a zero false positives approach, since it is impossible to distinguish between true positives and false positives otherwise.

\begin{figure*}[hptb]
    \centering
    \includegraphics[width=0.8\textwidth]{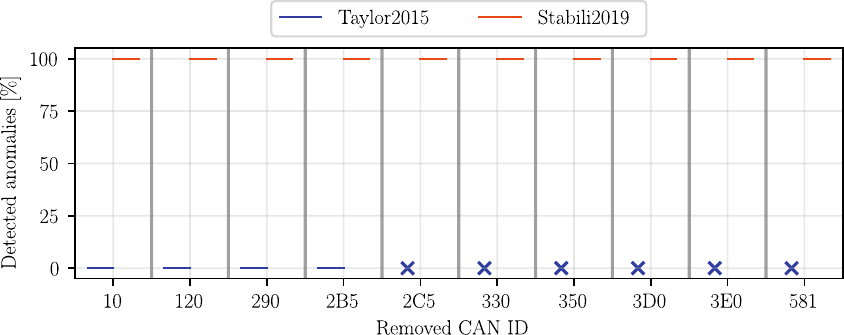}
    \caption{Experimental evaluation of the detection algorithms against the message removal attack scenario of the Ventus dataset}\label{fig:removal_results}
\end{figure*}

The detection results of the \Taylor and \Stabili against the message removal attack are presented in Figure~\ref{fig:removal_results}, where the percentage of detected anomalies ($y$-axis) for each removed message ID ($x$-axis) is shown.

The percentage of detected anomalies is presented using the box-plots. The detection performance of the two algorithms against this attack scenario are extremely consistent across the different missing IDs, hence it is nearly impossible to notice the different percentiles since they overlap with the median. We recall that \Taylor is designed to support only messages with a cycle time below $50$ milliseconds, hence it is possible to use it only against the removal of the first $4$ message IDs.
From the analysis of the results depicted in Figure~\ref{fig:removal_results} it is clear that our implementation of \Taylor is not able to reliably detect anomalies also against this attack scenario, since the removal of messages from normal CAN communications does not introduce significant deviations of the anomaly score used by \Taylor. 
The detection performance achieved by \Stabili are close to $100\%$, although this ideal value is never reached. This aspect has already been addressed in~\cite{Stabili2019Missing} and is related to the introduction of the \textit{valid waiting time} required to achieve zero false positives in the validation process. 

\subsection{Detection results against the OTIDS dataset}
\label{ss:detection_otids}
In this section we present the performance evaluation of the detection algorithms against the OTIDS dataset. To perform an evaluation of the detection algorithms, a labeled dataset is required, and since the OTIDS dataset is not labeled, we recreated the labels by following the description of the attacks. The three attack scenarios included in the OTIDS dataset are described as follows:
\begin{itemize}
    \item \textbf{Fuzzy attack:} the attack starts after $250$ seconds of normal traffic, and includes both normal and injected messages with $8$ different message IDs: $0x153$, $0x164$, $0x1F1$, $0x220$, $0x2C0$, $0x4B0$, $0x4B1$, and $0x5A0$. 
    \item \textbf{Denial-of-Service attack:} the attack starts at the beginning of the trace, and injects messages with ID $0x0$. 
    \item \textbf{Impersonation attack:} the attack starts after $250$ seconds of normal traffic, removes all messages with ID $0x164$ and inject messages with the same message ID mimicking its normal cycle time.
\end{itemize}
Following the attack description, we remark that it is not possible to distinguish between normal and injected messages in case of the \textit{fuzzy} attack scenario, hence we can not use that particular attack scenario in our experiments, with only the \textit{DoS} and the \textit{impersonation} attack scenarios available for performance evaluation. However, we remark also that in the \textit{DoS} attack scenario the injected message ID is not found in the training trace, thus making the detection task trivial by simply checking if there is a reference for the ID in the detection model. Hence, the only attack scenario from the OTIDS dataset that can be used for performance evaluation is the \textit{impersonation} attack.

\begin{figure*}[hptb]
    \centering
    \includegraphics[width=0.8\textwidth]{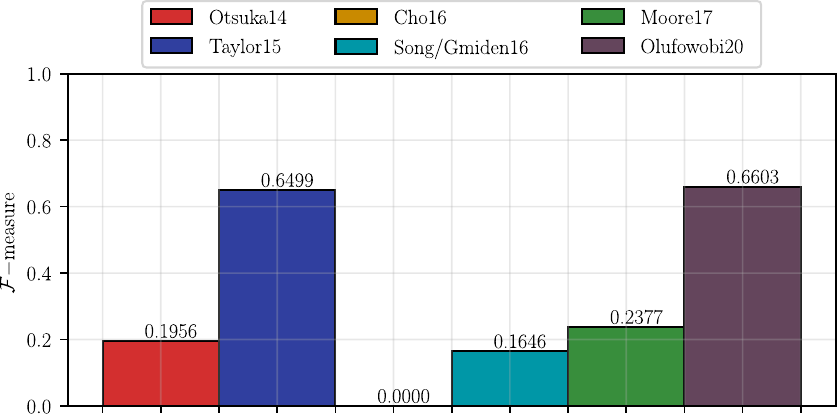}
    \caption{Experimental evaluation of the detection algorithms against the impersonation attack scenario of the OTIDS dataset}\label{fig:otids_impersonation}
\end{figure*}

The comparison of the performance evaluation of the algorithms against the impersonation attack scenario of the OTIDS dataset is shown in Figure~\ref{fig:otids_impersonation}. Figure~\ref{fig:otids_impersonation} shows, for each detection algorithm, the $\mathcal{F}-$measure evaluated on the impersonation attack scenario. 

The results shown in Figure~\ref{fig:otids_impersonation} shows that in this particular attack scenario the algorithm that is able to achieve the highest detection performance are \Taylor and \Olufowobi, while the other detection techniques struggles to achieve a $\mathcal{F}-$measure close to $0.5$, with \Cho failing to detect a single anomaly ($\mathcal{F}-$measure equal to $0$). However, we remark that these results are relative to a single test case, and that by changing the injected message or its injection frequency the overall detection results might vary significantly, as observed previously in Section~\ref{sss:infected_dataset}.

In the OTIDS impersonation attack scenario, the impersonated message ID $0x164$ is one of the most frequent messages, with a cycle time of $9$ milliseconds. The impersonation attack requires the removal of the target message after $250$ seconds, and the injection of a single message with the same message ID every $9$ milliseconds, mimicking the original frequency. In these conditions, we remark that the impersonation attack of the OTIDS dataset is actually a particular scenario of masquerading attack, in which the injected CAN messages substitute the normal ones.

However, following the performance evaluation of the tested algorithms we conducted manual analysis of the detection results to understand the behavior of the detection algorithms. The first interesting results from this analysis is that \Otsuka, \Song, and \Moore raise anomalies after the start of the attack. By analyzing the beginning of the attack simulation, we discovered that in the first $18$ milliseconds following the start of the attack there are $4$ different messages with ID $0x164$ instead of the expected $2$. This implies that in this time window the attack scenario is closer to the message injection attack considered in the threat model of Ventus, allowing \Otsuka, \Song, and \Moore to detect anomalies.
The second interesting result is that both \Taylor and \Olufowobi raised the majority of their alarms after approximately $50$ milliseconds since the start of the attack, being able to identify manipulated messages as anomalous. This implies that the evaluated detection performance of these two algorithms are to be considered as against the impersonation/masquerade attack scenario and not relative to the message injection attack. This interesting result might also be the cause for the low detection performance of both algorithms against a real message injection attack scenario as presented earlier in Section~\ref{sss:injection_comparison}. As a final remark, we highlight that the detection performance of \Cho are heavily affected by the values of its configuration parameter. However, we were not able to identify any combination of values that allows \Cho to identify anomalies against the attacks on the OTIDS dataset.

%% file: sections/conclusions.tex
\section{Conclusions}
\label{s:conclusions}

This paper contributes to the state-of-the-art by (I) surveying and implementing eight different detection algorithms based on CAN message timing analysis; by (II) publicly releasing their reference implementations; and by (III) testing the implemented algorithms against two different datasets, to present a detailed comparison of the detection performance of the analyzed detection algorithms against the same threat model and using the same detection metrics (\fmeasure and detection rate).
The novel dataset used for our experimental evaluation~\cite{Pollicino2021TCPS}, which is composed by more than $90$ minutes of training data and more than $400$ CAN traces containing different labeled attacks, is publicly available to advance current solutions.
With respect to the current state-of-the-art, this work presents an empirical and unbiased comparison of timing-based detection algorithms against the threat model of two different datasets, by addressing reproducibility of the results and highlighting the limitations of the dataset already publicly available for the performance evaluation of CAN anomaly detection algorithms. All the implemented algorithms and the dataset used in our experimental evaluation are publicly available to enable further improvements on this research topic.

Our main motivation is the impossibility of a direct comparison of similar proposals, due to inherent limitations of the literature. Authors usually do not release the source code of their implementations. Detection algorithms are described with an insufficient level of detail, thus requiring additional assumptions for their implementation that might have a considerable impact on detection performance. Different algorithms are tested on different private datasets, and the same attack is often implemented in different ways. These issues make it impossible to demonstrate an advancement over the state of the art, and even to compare novel proposals against it. This work solves all the aforementioned issues, thus establishing a fair, transparent and open-source baseline that can be used by all researchers and industry practitioners. 
